\documentclass[aps,prl,twocolumn,groupedaddress]{revtex4}

\usepackage{graphicx}
\usepackage{dcolumn}
\usepackage{bm}
\usepackage[T1]{fontenc}
\usepackage{color}

\begin{document}

\title{Edge Magneto-Fingerprints in Disordered Graphene Nanoribbons}
\author{Jean-Marie Poumirol$^1$, Alessandro Cresti$^2$, Stephan Roche$^{3,4}$, Walter Escoffier$^1$, Michel Goiran$^1$, Xinran Wang$^5$, Xiaolin Li$^5$, Hongjie Dai$^5$, Bertrand Raquet$^1$}
\affiliation{$^1$Laboratoire National des Champs Magn\'{e}tiques
Intenses, INSA UPS CNRS, UPR 3228, Universit\'{e} de Toulouse, 143
av. de Rangueil, 31400 Toulouse, France}
\affiliation{$^2$ CEA, LETI, MINATEC, F38054 Grenoble, France}
\affiliation{$^3$ CEA, INAC, SP2M, $L_{sim}$, 17 avenue des Martyrs, Grenoble, France}
\affiliation{$^4$ CIN2 (CSIC-ICN), Campus UAB, E-08193 Barcelona, Spain}
\affiliation{$^5$Department of Chemistry and Laboratory of
Advanced Materials, Stanford University Stanford, California
94305, USA}

\begin{abstract}
We report on (magneto)-transport experiments in chemically derived
narrow graphene nanoribbons under high magnetic fields (up to 60
Tesla). Evidences of field-dependent electronic confinement
features are given, and allow estimating the possible ribbon edge
symmetry. Besides, the measured large positive magnetoconductance
indicates a strong suppression of backscattering induced by the
magnetic field. Such scenario is supported by quantum simulations
which consider different types of underlying disorders (smooth
edge disorder and long range Coulomb scatters).
\end{abstract}

\pacs{72.80.Vp,75.47.-m,73.22.Pr}
%72.80.Vp   Electronic transport in graphene
%73.22.Pr   Electronic structure of graphene
%75.47.-m   Magnetotransport phenomena; materials for magnetotransport
%72.15.Rn   Localization effects (Anderson or weak localization)
\maketitle

{\it Introduction}.-The control of the current flow in graphene
nanoribbons (GNRs) constitutes a fascinating challenge for the
future of carbon-based electronic devices. The description of low
energy excitations as massless Dirac fermions in graphene
(resulting in large mobility at room temperature \cite{Geim07}),
together with the possibility for bandgap engineering \cite{Han07}
suggest a strategy to outperform Silicon devices
\cite{Liang07,Wang08}. However, the lateral size reduction goes
along with a problematic decay of the charge mobility
\cite{Wang08,Lin08}. Low temperature transport measurements
performed at zero magnetic field on lithographically patterned
GNRs unveil an ubiquitous energy gap, irrespective of the edges
symmetry \cite{Han07}, exceeding the expected confinement gap
\cite{Son06,Stampfer09} and driven by disorder induced charging
effects \cite{Han10, Gallagher10}. Non-perfect edges are also
expected to produce a significant scattering source. Drastic
consequences on the electronic transport have been theoretical
anticipated with the formation of a mobility gap, even in the
presence of an ultra-smooth edge roughness \cite{Areshkin07}.
Other sources of disorder like bulk vacancies \cite{Ihnatsenka09},
charge trapped in the oxide \cite{Stampfer09,Gallagher10}, or
structural deformations (ripples) \cite{ripples} are also believed
to significantly alter the conductance, although the dominant
scattering source remains debated, and seems to be sample
dependent. In that perspective, experimental works on narrow GNRs
remain sorely lacking.

This Letter presents compelling evidences of the 1D transport
character in a 11nm wide GNR, and the possibility of tuning
backscattering effects by means of an external magnetic field.
Bandstructure calculations allow some assignment of the measured
gate-dependent conductance modulations to the underlying van Hove
singularities, and hence some estimation of the likely ribbon edge
symmetry. The application of perpendicular high magnetic field
further induces a marked enhancement of the conductance,
irrespective of the applied gate voltage and in large contrast to
the magnetofingerprints of graphene flakes \cite{Novoselov06}.
Close to the charge neutrality point, the measured large positive
magnetoconductance (MC) is attributed to the formation of the
first Landau state, responsible for the closing of the energy gap
and of a marked reduction of backscattering processes.
Landauer-B\"uttiker conductance simulations convincingly support
the scenario of an entangled interplay between the magnetic bands
formation and a disorder-induced interband scattering suppression.
Both edge disorder and long range coulomb scatters yield similar
conclusions.

{\it Transport measurements}.-Experiments are carried out on the
first generation of chemically derived GNRs with smooth edges
 \cite{Li08b} deposited on ${\rm Si/SiO_2(500nm)}$ wafer and
connected to Pd-electrodes distanced by 270nm. The electrostatic
doping is monitored by a back-gate voltage. In the following, we
focus on a 11nm wide GNR for which the first Landau orbit of
radius $l_B=\sqrt{\hbar/eB}$ reaches the ribbon width of our GNR
sample at 30T. Complementary measurements on larger GNR, here
90nm, are briefly discussed to emphasize the effect of the
electronic confinement on the magnetoconductance.

\begin{figure} [!ht]
\center
\includegraphics[width=8 cm]{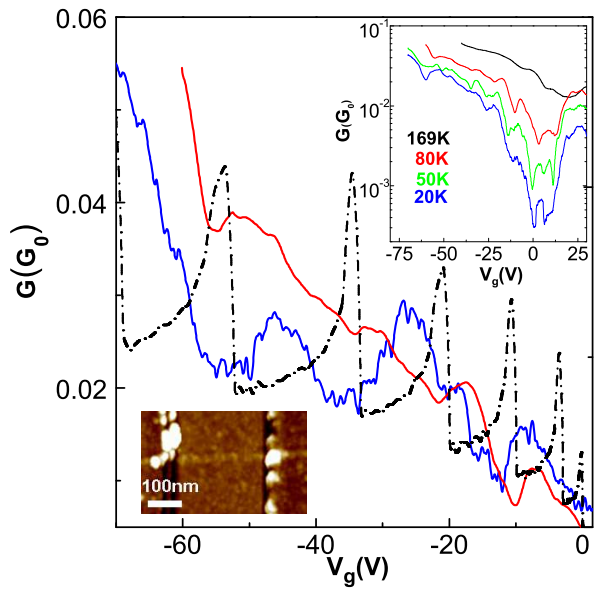}
\caption{(color online) Conductance versus $V_g$ at 80K measured
on the narrow GNR (width $\simeq 11$nm) and for two $V_{b}$, 50mV
and 1mV (respectively red and blue curves). The density of states
of a 90-aGNR is superimposed (dashed curve). Top inset:  $G(V_g)$
at various temperatures for a wider gate voltage range. Bottom
inset: AFM image of the measured GNR device.}
\end{figure}

Figure 1 (inset) shows the gate voltage dependent conductance of
the 11nm wide GNR under 50mV bias voltage ($V_b$), and at various
temperatures (from 169K down to 20K). The V-shape curves are
typical for a semiconducting ribbon \cite{Han10}. The conductance
at the CNP measured at $V_b={\rm 1mV}$ follows a thermally
activated behaviour down to 50K. Interestingly, a detailed
analysis of the $G(V_g)$ curves at 80K and below unveils
reproducible modulations superimposed to the overall increase of
the conductance versus gate voltage (Fig.1, main frame). These
structures evolve to pronounced conductance drops for low bias
measurements (blue curve obtained at 1mV). We assign such
conductance profile to the presence of van Hove singularities
(vHs), responsible for the enhancement of the backscattering in
the diffusive regime. To validate such scenario, extensive
tight-binding bandstructures calculations are performed on a set
of ribbons symmetries with varying widths \cite{Alessandro}.
Calculations include zigzag (zGNR) and armchair GNR (aGNR) of
three types N=3m, 3m+1 and 3m+2, N being the number of dimer
lines. The N values are chosen to explore the full range of the
ribbon sample width dispersion (here $11\pm1.5{\rm nm}$), as
evaluated by careful AFM analysis. Figure 1 also shows the density
of states $\rho(V_g)$ (dotted-dashed curve) for the armchair
configuration of type 3m with N = 90 corresponding to a nominal
width of 10.947nm, which turns out to be the most likely ribbon
geometry obtained by the theoretical analysis \cite{Alessandro}.
Indeed, without any free parameters, a convincing agreement is
observed between the local minima of conductance and the
theoretical sequence of vHs's onsets. Such correspondence is not
achieved at all for zGNR and other 3m+1 and 3m+2 types of aGNR
\cite{Alessandro}. Note that other N values, multiple of 3, from
87 to 93, well below the experimental width uncertainty, also lead
to a satisfactory agreement. The bands arrangement undergoes a
scaling of the order of 6\%, which cannot be discriminated on the
$G(V_g)$ curves. We conclude that the conductance exhibits 1D
subbands fingerprints consistent with an armchair arrangement of
the carbon atoms at the edges with 3m dimer lines, around
$90\pm3$, along its width.

\begin{figure} [!ht]
\center \resizebox{\columnwidth}{!}{\includegraphics{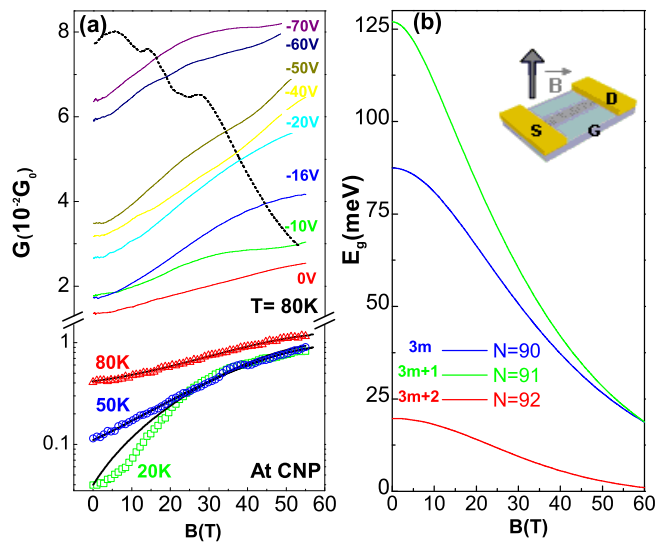}}
\caption{(color online) (a) Left panel: Magnetoconductance at 80K
for the narrow ribbon, and various gate voltages (top); and at
80K, 50K, 20K for $V_g=0$ (curves with symbols: measurements;
solid lines : simulated data). The MC for a larger width GNR
(90nm) is also shown (dashed curve). (b) Right panel: Computed
field-dependent energy gaps for three different types of aGNR.
Inset: Schematic of the GNR device.}
\end{figure}

{\it Magnetoconductance behavior}.-The application of a transverse
magnetic field on the 11nm wide GNR drastically increases its
conductance by more than 50\% at 80K and by more than one decade
at 20K (Fig.2-left panel). The gain of transmission is further
modulated by the electrostatic doping in a complex but very
reproducible manner. One notes that the MC remains of large
amplitude, even when several subbands are involved in conduction.
Conversely, for a 90nm wide GNR (Fig.2a, black dashed curve) and
generally speaking for graphene flakes \cite{Tikhonenko08}, only
the very low magnetic field MC is slightly positive and results
from a weak localization regime. At larger field, the conductance
of large width GNR is strongly reduced, following the two-fluid
model at the CNP in a diffusive regime \cite{Cho08} and further
exhibits a step-like decrease in the quantum regime, unveiling the
graphene Landau Levels (LL) \cite{Gusynin05} and the reduction of
the number of available channels at the edges \cite{Li09}. In the
following, we bring evidence that the large \emph{positive} MC and
its energy dependence obtained for the narrow GNR result from a
subtle interplay between the specific magnetic bandstructures and
the field induced reduction of disorder-driven backscattering.

The magnetic field dependent bandstructure calculations reveal
that the 1D subbands of the GNR evolve to the graphene LL above
30T, once the magnetic confinement gradually overcomes the
electronic one \cite{Alessandro}. At the CNP, the onset of the
zero-energy LL induces a closing of the energy bandgap of the GNR.
In Fig.2b the simulated magnetic field dependence of the energy
gaps for the three types of aGNR are reported. Assuming a flat
band model at the CNP, the magnetic-field dependence of the ribbon
energy gaps $E_g(B)$ will modulate the thermally activated regime,
in which the magnetoconductance is expressed as $\Delta
G(B)\propto {\rm exp}(-\frac{E_g(B)}{2k(T+T^*)})$. Here, $kT^*$
accounts the kinetic energy window of charge carriers defined by
$eV_{b}/2$.  As seen in Fig.2 (left panel), an excellent agreement
between experiments and simulation is obtained for the 90-aGNR
energy gap with $kT^*\approx 21\pm 2{\rm meV}$, which is
consistent with the 50mV experimental $V_{b}$.  We thus conclude
that if the sequence of the vHs for an aGNR of type 3m appears as
the most likely fingerprint of the $G(V_g)$ modulations at zero
field, the agreement is further reinforced by the large positive
MC.

{\it Transport simulations}.-To deepen the understanding of the
large positive magnetoconductance signal, we perform tight-binding
transport simulations using the Green functions formalism
\cite{Cresti03}. Three different types of disorders are
considered: namely the Anderson disorder (with on-site energies
varying at random in the range $[-w/2,w/2]$ and $w=2$ eV); the
edge roughness with a 5\% of removed carbon atoms on the six outer
chains at each edge; and randomly dispersed impurities described
by a long range Gaussian potential with a density $1.15\times
10^{16}$ m$^{-2}$ (spatial range $\xi=1$ nm, maximum strength in
the range $[-u/2,u/2]$ with $u=2$ eV). These disorders are well
representative of the possible scattering sources considered in
the literature \cite{Areshkin07, Ihnatsenka09,ripples}. In
particular, edge roughness seems an unavoidable feature of
nanoribbons, even when chemically derived; whereas impurities with
long-range potential mimic screened charged ions trapped in the
oxide. Finally Anderson disorder qualitatively reproduces the
presence of impurities on the surface of the GNR. Transport
calculations are performed at finite temperature T=80 K and for a
drain-source bias $V_b=50 $ mV. The contribution of a contact
resistance of 10 k$\Omega$ in the ON-state of the current flow is
also included \cite{Wang08}.

\begin{figure}
\center \resizebox{\columnwidth}{!}{\includegraphics{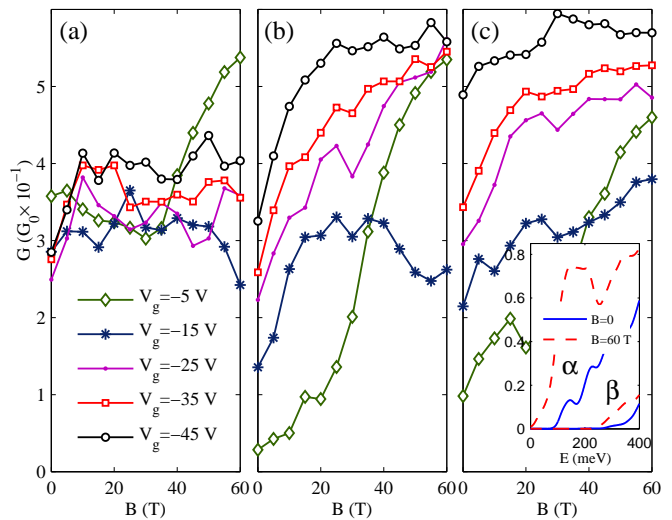}}
\caption{\label{fig3}(color online) Magnetoconductance for the
90-aGNR at different $V_g$ with (a) Anderson disorder, (b) edge
roughness and (c) Gaussian impurities. Inset of (c): $\alpha$ and
$\beta$ transmission factors denote the contributions of the first
channel and the higher bands at zero field and 60 T.}
\end{figure}

\begin{figure*}
\center \resizebox{\textwidth}{!}{\includegraphics{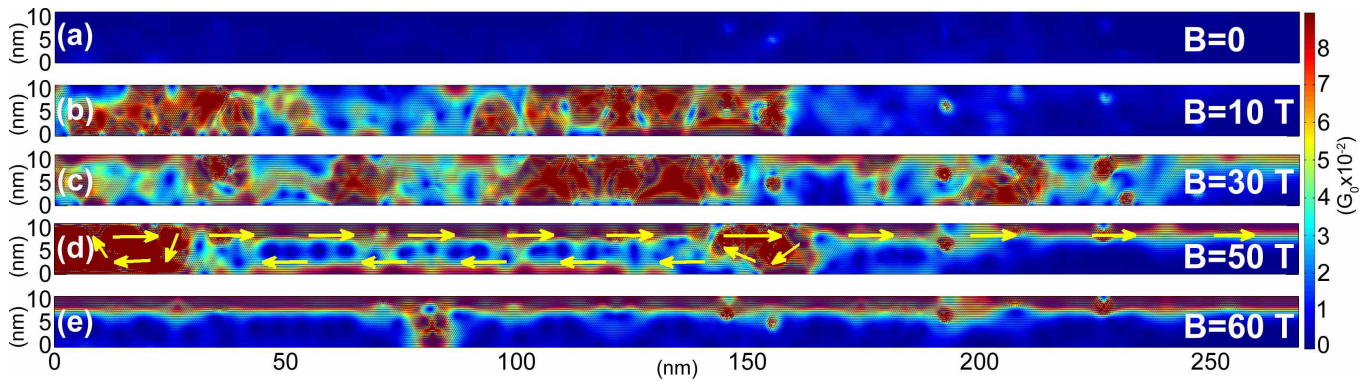}}
\caption{\label{fig4}(color online) Spectral current distribution
at the energy $E=200$meV for a 90-aGNR in the presence of Gaussian
disorder only at (a) $B$=0, (b) $B$=10 T, (c) $B$=30 T, (d) $B$=50
T, (e) $B$=60 T. Charge flows from left to right. }
\end{figure*}

Figure \ref{fig3} shows the computed magnetoconductance profiles
(at several gate voltages) for each type of disorder. First, it is
clear that the use of the Anderson disorder seems inappropriate to
describe the experimental magnetofingerprints. The conductance is
indeed weakly sensitive to the magnetic field (only effective at
low $V_g\leq -5$ V), and this behavior hardly changes with further
modulation of the disorder strength ($w$) (not shown here). In
contrast, a huge positive MC is observed at low $V_g$ (where only
a single conductive channel is active) for the case of edge
roughness. When two or three channels are available for conduction
($-15{\rm V}< V_g \leq -5{\rm V}$), the MC displays however a more
fluctuating trend, but becomes again positive at higher gate
voltages (see $V_g=-25$V). This partially reproduces the general
trend of the measured MC, suggesting that edge roughness can
partly account for disorder effects but does not fully dominate
the scattering processes. Finally, Gaussian disorder qualitatively
reproduces the experimental trend for all $V_g$. As in
experiments, the conductance increases with gate voltage and the
MC is positive at almost any magnetic field and $V_g$ values. A
more quantitative agreement between simulated and measured
conductance data can only be achieved by considering a Gaussian
disorder superimposed to an ultra-smooth roughness at the edges
\cite{Alessandro}.

The positive MC behavior obtained for Gaussian impurities (or edge
disorder) can be further rationalized by discussing and
visualizing the magnetic edge states that develop in the GNR in
the high magnetic field limit. These field-induced edge states
have been already discussed in the context of quantum wires
\cite{Cresti08b}. Due to the broken time-reversal symmetry, edge
channels on the upper edge are created and convey current in
opposite direction than the channels at the lower edge. If
disorder is not strong enough, the spatial separation between the
channels induced by the high magnetic field provokes a strong
suppression of backscattering (and positive magnetoconductance).
Figure 4 images the spectral current distribution at the energy
$E=200$meV for several magnetic fields. At zero magnetic field
(a), the current flow is weak (as well as related conductance).
Increasing $B$ magnifies the current density along the ribbon
(Fig.4-b,c), but the presence of scatters limit the total current
flow between source and drain. Figure 4(d) illustrates the partial
backscattering along the lower edge (where electrons can only move
from right to left) and partial transmission to the right contact
along the upper edge (yellow arrows), when the current flow
reaches a localized state. This effect is further reduced with the
formation of fully developped magnetic edges at higher magnetic
field (see (e) for $B=60$T) which further enhances the positive MC
signal.

One notes that for the Anderson disorder, the homogeneous
short-range scattering potential makes strongly difficult a
spatial separation between the chiral channels, thus prohibiting
the predominance of magnetic edge channels. In contrast, the
mechanism is efficient on the first conductive channel when bulk
disorder is absent (edge roughness) or weak and long-range
scattering potential prevails. In this case, intraband scattering
(within the first energy band) is partially suppressed and the
conductance increases as the magnetic channels are pushed to the
edges. A detailed analysis disentangles the contribution of the
first channel from the higher subbands (represented by the
$\alpha$ and $\beta$ factors, respectively \cite{Alessandro}). For
a Gaussian disorder, $\alpha\gg \beta$ (Fig.3c-inset), so the
first channel dominates transport and its magnetic field
dependence. The long-range nature of the Gaussian potential
markedly reduces the interband scattering and preserves the
positive MC effect of the first channel well beyond the second
vHs, as also observed experimentally. The higher channels are less
active and do not show any marked positive MC. This is explained
by the spatial enlargement of the magnetic edge states, which then
come into contact more easily, due to the narrow width of the
ribbon. Conversely, the short range nature of edge disorder
favours enhanced interband scattering and mixing between channels,
as evident from Fig.\ref{fig3}b, where the positive MC is reduced
when the second and third subbands start to be involved at $V_g
\simeq -10$ V.

{\it Conclusion}.-In large magnetic field, charge transport
properties in chemically derived graphene nanoribbons can be tuned
by the formation of magnetic edge states. The resulting positive
magnetoconductance fingerprint obtained experimentally has been
further satisfactorily reproduced by quantum simulations,
pinpointing the likely contribution of both charged impurities and
smooth edge roughness.

Part of this work has been supported by EuroMagNET, EU contract
n$^o$ 228043. S.R. acknowledges the ANR/P3N2009 (project ${\rm
NANOSIM{\_}GRAPHENE}$ n$^o$ ANR-09-NANO-016-01). H.D acknowledges
funding from Office of Naval Research (ONR) and Intel.

\end{document}